\documentclass[aps,reprint,superscriptaddress,showkeys]{revtex4-2}
\usepackage{graphicx}
\usepackage{dcolumn}
\usepackage{bm}
\usepackage[mathlines]{lineno}
\usepackage{epsfig}
\usepackage{epstopdf}
\usepackage{xcolor}
\usepackage{mathtools}
\usepackage{subfig}

\begin{document}
\title{Predicting the turbulent transport of cosmic rays via neural networks}
\author{D. I. Palade}
\email{dragos.palade@inflpr.ro}
\affiliation{National Institute of Laser, Plasma and Radiation Physics,	M\u{a}gurele, Bucharest, Romania}%
\affiliation{Faculty of Physics, University of Bucharest, Măgurele, Romania}
\date{\today}

\begin{abstract}
A fast artificial neural network is developed for the prediction of cosmic ray transport in turbulent astrophysical magnetic fields. The setup is trained and tested on bespoke datasets that are constructed with the aid of test-particle numerical simulations of relativistic cosmic ray dynamics in synthetic stochastic fields. The neural network uses, as input, particle and field properties and estimates transport coefficients $10^7$ faster than standard numerical simulations with an overall error of $\sim 5 \%$ . 
\end{abstract}

\keywords{cosmic ray, magnetic turbulence, turbulent transport, neural networks}

\maketitle

\section{Introduction}
\label{Section_0}


Cosmic rays (CRs) are charged particles originating mostly in galactic supernovae remnants and being accelerated by the Fermi mechanism in shock waves \cite{DRURY201252,Blasi2013} although, more exotic sources are possible \cite{Gabici}. Due to their (ultra)relativistic energies, CRs can permeate systems of all sizes, from the heliosphere, to galaxies and even the extra-galactic medium \cite{Gabici, doi:10.1146/annurev-astro-081710-102620}. Consequently, CRs carry important information to our understanding of fundamental high-energy physics, astrophysical magnetic fields, the structure of astrophysical media, space weather, etc. \cite{MAVROMICHALAKI20061141,Dvornikov2013,BHATTACHARJEE2000109}. 

The dynamics of CRs is a long-standing fundamental problem of astrophysics with many open questions that are difficult to solve for several reasons. During their journey from sources to detection, CRs interact mainly with magnetic fields. The latter are highly turbulent \cite{Schekochihin2007} and, via non-linear interactions with particles, lead to consistent (anomalous) transport phenomena beyond ballistic motion \cite{Giacalone_1999}. The CR's energies cover a wide range of magnitudes (from $MeV$ to $10^{20}eV$ \cite{Blasi2013}) while the turbulent features of the magnetic fields are quite diverse, spanning different space-scales, anisotropies, and fluctuation amplitudes \cite{BRANDENBURG20051}. This richness in physical regimes opens a variety of possible transport types ranging from subdiffusive to superdiffusive in the perpendicular and parallel directions, with complicated dependencies \cite{Giacalone_1999,BRANDENBURG20051,Xu_2013,PhysRevD.65.023002}.

Despite this discouraging picture, a lot of progress has been made in the past decades in the topic of turbulent CRs transport. Quasilinear approaches \cite{jokipii1966cosmic} and non-linear extensions \cite{Shalchi} gained a lot of attention due to their technical simplicity and ability to provide insight in the physical mechanisms involved. Unfortunately, such models are known to fail in a lot of relevant cases such as the limit of strong turbulence or the \emph{$90$ degree problem}. A much more accurate approach, which gained momentum in the scientific community, is the method of test-particle simulations either in synthetic \cite{PhysRevD.102.103016,POMMOIS2005647,Mertsch2020,Hauff_2010} or MHD generated \cite{Maiti_2022,Beresnyak_2011} turbulent magnetic fields. Within this approach, the dynamics of CRs is mimicked at the statistical level with an ensemble of fictitious particles that allows for a direct evaluation of the transport coefficients. Its main drawback is the relatively high numerical effort required, which is hardly compatible with the diversity of possible physical regimes. It is clear that the astrophysics community would benefit from a fast and accurate method of prediction in their quest to understand the CR turbulent transport.
 
A helping hand might come from another scientific front. In the recent years, we have witnessed the rise of artificial intelligence (AI) methods, in particular machine learning (ML) techniques \cite{doi:10.1126/science.aaa8415,ABIODUN2018e00938,golden1996mathematical}, that are able to provide inferences in various mathematical problems. From the multitude of ML methods, the reader is directed towards artificial neural networks (ANNs) \cite{golden1996mathematical,ABIODUN2018e00938}. The promise of ML is that, regardless of the chosen technique, if sufficient data is available, it can learn from that data and make fast and reliable prognoses for unknown cases. Such abilities would be equivalent to having analytical expressions at hand and would bypass the technical difficulties that arise in doing many simulations of CR dynamics. 

Given this scientific context, the purpose of the present work is to illustrate the methodology for developing an artificial neural network designed for predictions of cosmic ray turbulent transport in astrophysical magnetic fields. For the training and testing phases, a database is constructed with the aid of test-particle numerical simulations in synthetically generated fields. Since the main purpose is methodological, we restrict ourselves here to the evaluation of perpendicular diffusion \cite{Giacalone_1999} in a relatively limited range of physical parameters. Nonetheless, it is the hope of the author that this work will open a path towards more extensive databases and, consequently, more potent ANNs.

The rest of this manuscript is organized as follows. The Theory section \ref{Section_1} describes, briefly, the CR transport model \eqref{Section_1.1}, the test-particle numerical method \eqref{Section_1.2}, the general architecture of ANNs \eqref{Section_1.3} and the construction of a training and testing database \eqref{Section_1.4}. The Results Section \eqref{Section_2} presents the convergence properties and the predictive power of the ANN. In the Conclusions section \eqref{Section_3}, the main findings are resumed and perspectives are discussed.

\section{Theory}
\label{Section_1}

\begin{figure}
	\includegraphics[width=.98\linewidth]{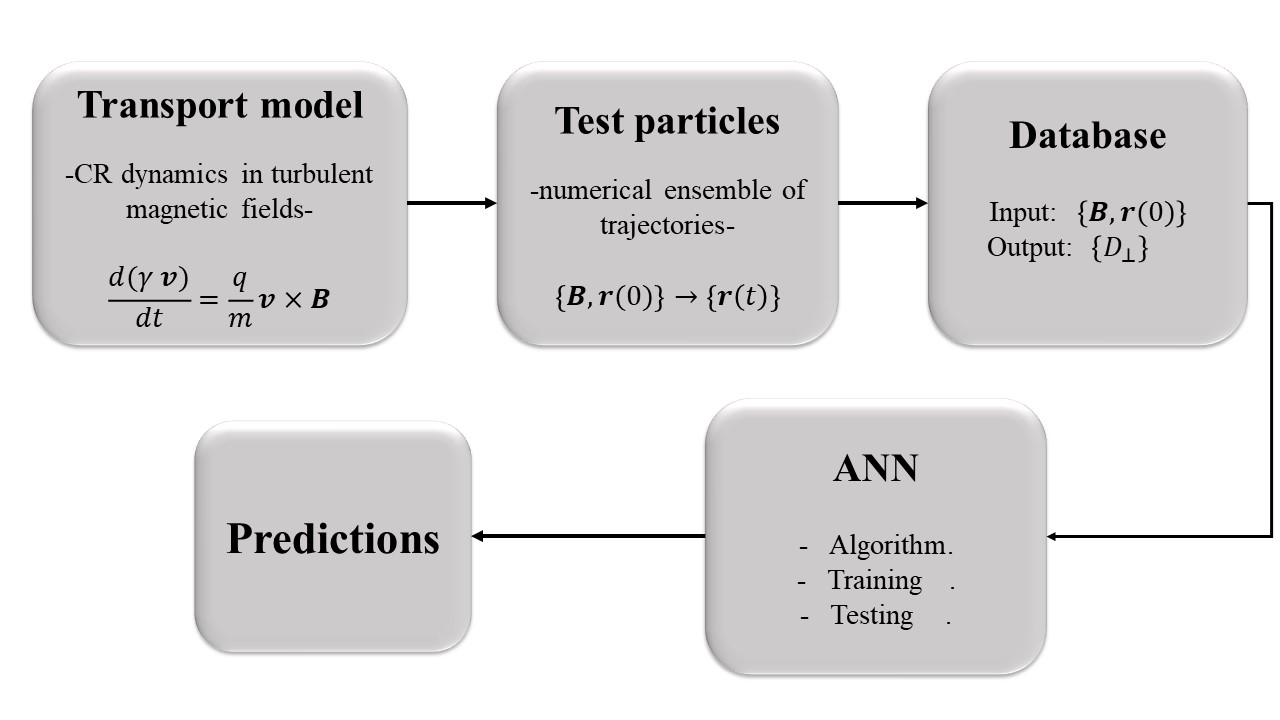}%
	\caption{Schematic view of the elements required to construct the ANN.}
	\label{fig_00}
\end{figure}

The ingredients of an ANN are the programming structure and the training dataset. For our problem, the latter is constituted of input-output pairs representing field-particle properties and diffusion coefficients. Such data is obtained with the numerical method of test particles that move according to a transport model. In this section, these elements, shown schematically in Fig. \eqref{fig_00}, are discussed in reverse order.

\subsection{The transport model}
\label{Section_1.1}

A cosmic ray is a charged particle characterized by its rest mass $m$, charge $q$, position $\mathbf{r}(t)$ and velocity $\mathbf{v}(t)=d\mathbf{r}(t)/dt$ in a global reference frame obeying the relativistic Newton-Lorentz equation in the presence of an astrophysical magnetic field $\mathbf{B}$:

\begin{align}\label{eq_1.1}
	\frac{d(\gamma \mathbf{v})}{dt}&=\frac{q}{m}\left(\mathbf{v}\times\mathbf{B}\right).
	\end{align}

The kinetic energy of the particle is $T=mc^2(\gamma-1)$ where the Lorentz factor $\gamma=(1-v^2/c^2)^{-1/2}$, $c$ is the speed of light and any electric field is neglected $\mathbf{E}=0$. A cartesian system of coordinates $\mathbf{r}\equiv (x,y,z)$ is defined. The magnetic field $\mathbf{B}$ is decomposable into an average constant component $B_0$ along $Oz$ and a fluctuating part $\mathbf{b}$ which is turbulent. Relative to this decomposition we coin the $Oz$ direction as being "parallel" $\hat{e}_z\equiv \hat{e}_\parallel$ and the plane $(x,y)$ as "perpendicular". Naturally, for any wavevector  $\mathbf{k}=(k_x,k_y,k_z)$, $k_z\equiv k_\parallel$ and $(k_x,k_y)\equiv \mathbf{k}_\perp$. The velocity $\mathbf{v} = v_\parallel \hat{e}_\parallel+\mathbf{v}_\perp$ can be used to define the pitch angle $\mu = v_\parallel/v$, where $v^2=|\mathbf{v}|^2$. 

The turbulent component $\mathbf{b}$ of the total magnetostatic field $\mathbf{B} = B_0\hat{e}_\parallel+\mathbf{b}$ is represented in the paradigm of a 2D model \cite{POMMOIS2005647,Hauff_2010,Hussein}: 

\begin{align}\label{eq_1.2}
	\mathbf{b}(\mathbf{r}) &= \hat{e}_\parallel\times \nabla_\perp  a_\parallel(\mathbf{r}).
\end{align}

The magnetic vector potential $a_\parallel$ is Gaussian, zero-averaged $\langle a_\parallel(\mathbf{r})\rangle = 0$ and homogeneous \cite{Hussein,Mertsch2020,PhysRevD.102.103016}. The last property is evident in the spectrum of its fluctuations:

\begin{align}\label{eq_1.3}
	\langle \tilde{a}_\parallel(\mathbf{k}) \tilde{a}_\parallel(\mathbf{k}^\prime)\rangle &= S(\mathbf{k})\delta(\mathbf{k}+\mathbf{k}^\prime).
\end{align}

By $\tilde{a}_\parallel(\mathbf{k})$ was denoted the Fourier transform of $a_\parallel(\mathbf{r})$, while $\langle\cdot\rangle$ stands for statistical average. The overall turbulence amplitude is defined as $ b = \sqrt{\langle\mathbf{b}^2(\mathbf{r})\rangle}$ and the spectrum is assumed to be of Kolmogorov type with parallel-perpendicular anisotropic dependencies \cite{Hussein,Hauff_2010,PhysRevD.102.103016}: 

\begin{align}\label{eq_1.4}
	S(\mathbf{k}_\perp,k_\parallel) &\sim \frac{(k_\perp\lambda_\perp)^q\left(1+(k_\parallel\lambda_\parallel)^2\right)^{-s/2}}{k_\perp(1+(k_\perp\lambda_\perp)^2)^{(s+q)/2}},
	\end{align}
where $q=3, s=5/3$ and $\lambda_\parallel,\lambda_\perp$ are "bend-over" scales, intimately related to coherence/integral scales \cite{PhysRevD.102.103016}.

Within this transport model, realizations of the turbulent field $ \mathbf{b}$ drive associated CR trajectories $\mathbf{r}(t)$ via the eq. of motion \eqref{eq_1.1}. The ensemble of trajectories can be used to derive, as statistical averages, the diffusion transport coefficients in any direction $Ox$: 

\begin{align}\label{eq_1.5}
	D_{x,x}(t) = \frac{\langle \left(x(t)-x(0)\right)^2\rangle}{2t}.
\end{align}

In this paper, we are discussing regimes that are strictly diffusive and for which the quantity of interest is the asymptotic perpendicular diffusion coefficient $D_\perp = \lim_{t\to\infty}D_{x,x}(t)$. In this context, the problem of CR turbulent transport is equivalent to asking: \emph{what are the diffusion coefficients $Y=D_\perp$ for any given set of particle-field input parameters $X  = (T,\mu,b,\lambda_\perp,\lambda_\parallel)$}?

For practical reasons related to numerical implementation, all quantities involved in the transport model (eqns. \eqref{eq_1.1}-\eqref{eq_1.4}) are scaled as follows. The magnetic fields $\mathbf{b}\to \mathbf{b}B_0$, the time $t\to t\omega_c^{-1}$, the velocities $\mathbf{v}\to \mathbf{v}c$, the kinetic energy $T\to T mc^2 $, the space scales $(\mathbf{r},\lambda_i)\to (\mathbf{r},\lambda_i)\rho_L^0$ and the wave-vectors $\mathbf{k}\to\mathbf{k}/\rho_L^0$. The cyclotron frequency is defined as $\omega_c=|q|B_0/m$ while the "bare" Larmor radius $\rho_L^0=mc/(|q|B_0)$. Consequently, the diffusion coefficients are scaled as $D\to D m_0c^2/(qB_0)$.

\subsection{Test-particle numerical simulations}
\label{Section_1.2}

The present work employs the method of test-particle simulations \cite{PhysRevD.102.103016,Mertsch2020} (or direct numerical simulations \cite{Vlad_2021,Palade2021,Hauff_2010}) for the calculus of diffusion coefficients. The main idea is that we can mimic the reality of turbulent dynamics through a numerical representation of the transport model described in the previous subsection \eqref{Section_1.1}. A finite ensemble of $N_p$ random fields $\{\mathbf{b}\}$ with appropriate spectral properties (distribution and spectra) is constructed. For each realization of this ensemble, a CR trajectory $\mathbf{r}(t)$ can be computed by solving eq. \eqref{eq_1.1}. In the limit of many particles $N_p\to \infty$, the transport coefficients are given by statistical averages over the numerical ensemble of trajectories $\{\mathbf{r}(t)\}$ \eqref{eq_1.5}. More details on the method can be found in \cite{Palade2021}. 

The magnetic vector potential, $a_\parallel$, is assumed Gaussian, zero-averaged, and homogeneous; thus, it can be constructed numerically in each realization with a Fourier-like (harmonic) decomposition \cite{Palade_2023,Palade2021,Mertsch2020}:

\begin{align}\label{eq_1.6}
	 a_\parallel(\mathbf{r}) &= b\lambda_\perp\sqrt{\frac{2}{N_c}}\sum_{j=1}^{N_c} \sin(\mathbf{k}^j\mathbf{r} + \alpha^j)
\end{align}
where $\alpha^j$ are independent random initial phases distributed uniformly in the interval $[0,2\pi)$. The partial waves $\mathbf{k}^j\equiv\left(\mathbf{k}_\perp^j,k_\parallel^j\right)$ are independent random variables generated with the use of the acceptance-rejection algorithm \cite{accept} for the PDF $S(\mathbf{k})$. It can be shown that the representation \eqref{eq_1.6} is, indeed, zero-averaged with the correct spectrum \eqref{eq_1.4}. In the limit of many partial waves $N_c\to \infty$, the Gaussianity is also achieved (provided by the \emph{central limit theorem}). 

Solving eq. \eqref{eq_1.1} is a standard numerical problem which is tackled here with a relativistic Boris pusher \cite{10.1063/1.2837054}. Magnetic field values are evaluated in each realization directly using eqn. \eqref{eq_1.6}. Within the present scaling, the gyro-period is $2\pi$ and the time step is chosen $\delta t = 2\pi/10$. This level of time discretization was found to be sufficient for an accurate prediction of the Larmor rotation and CR dynamics. 

Practical experience has shown that using $N_p = 10^4$ and $N_c=500$ in a simulation is enough to achieve acceptable Gaussianity, Eulerian, and Lagrangian convergence, thus reliable results on transport.

\subsection{Artificial neural networks}
\label{Section_1.3}

\begin{figure}
	\includegraphics[width=.98\linewidth]{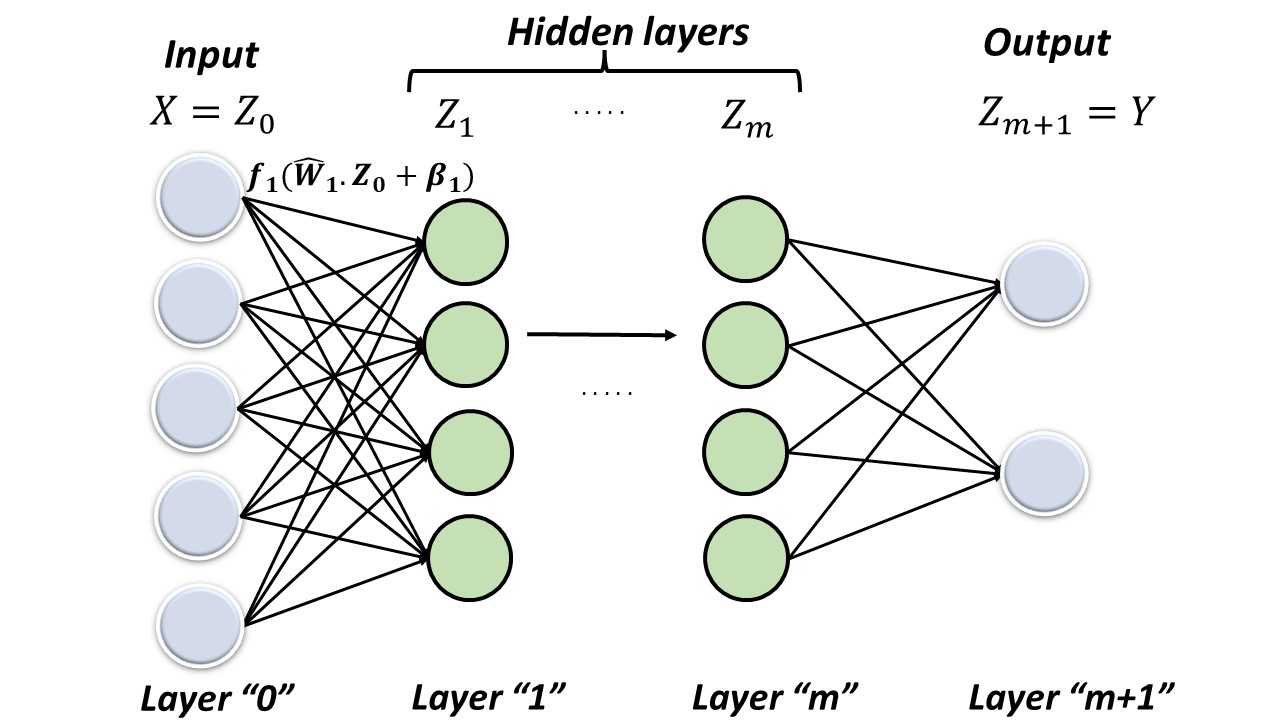}%
	\caption{The architecture of a general ANN with $5$ input and $2$ output variables.}
	\label{fig_0}
\end{figure}

In this section, the generic features of an ANN architecture that will be used for predicting the transport of CRs in turbulent magnetic media are described. Its structure is represented graphically in Fig. \eqref{fig_0}. For more details, see \cite{priddy2005artificial}. 

A general ANN is a computing setup consisting of $m+2$ "layers", each layer $i$ contains $n_i$ "neurons" and each neuron $j$ is described by a numerical value $Z_{i,j}$, a "bias" $\beta_{i,j}$, an "activation function" $f_{i,j}$ and some "weights" $W_{i,j,k}$. The latter connects recursively neurons from neighbouring layers, thus, $W$ is an irregular tensor of $(m+1)\times n_i\times n_{i-1}$ dimension. The first layer, $i=0$, contains the values of the input variable $X$, $Z_{0,j} = X(j)$. The last layer, $i=m+1$ contains the values of the output variable $Y$, $Z_{m+1,j} = Y(j)$. The rest are coined "hidden layers". Within an ANN there is a feed-forward recurrent relation between layers:

\begin{align}	\label{eq_1.7}
	Z_{i,j} = f_{i,j}\left(\sum_{k=1}^{n_{i-1}}W_{i,j,k}.Z_{i-1,k}+\beta_{i,j}\right)
\end{align}

The purpose of an ANN is to model the true relation (which is unknown) between the input variable $X$ and the output $Y$. This is achieved by finding appropriate weights and biases $\{W_{i,j,k},\beta_{i,j}\}$ that minimize an error function $E$ between the output of the network, $Z_{m+1}$, and the real output, $Y$. This must be done over a dataset of $N_d$ pairs $\{X^p,Y^p\}_{p=1,N_d}$ which should be \emph{reasonably large}. The error function $E$ is defined as Minkowski error: 

\begin{align}\label{eq_1.8}
	E = \frac{1}{n}\sum_{p=1}^{N_d}\sum_{j=1}^{n_{m+1}}\left(Z_{m+1,j}^p-Y^p(j)\right)^2 .
\end{align}

The error minimization is reached iteratively, over multiple "epochs", using ADAM \cite{kingma2014adam}, a modified version of gradient descent algorithm \cite{ruder2016overview}. In this approach, during each iteration (epoch), the weights and biases $\{W_{i,j,k}, \beta_{i,j}\}$ (denoted generically $\theta$), are updated using a "learning"/"training" procedure:
\begin{align}\label{eq_1.9a}
	v_\theta&= \alpha_1 v_\theta + (1-\alpha_1)\left(\frac{\partial E}{\partial \theta}\right)^2\\
	m_\theta&=\frac{\alpha_2 m_\theta }{1-\alpha_1^{epoch}}+\frac{1-\alpha_2}{1-\alpha_1^{epoch}}\left(\frac{\partial E}{\partial \theta}\right)\label{eq_1.9b}\\
	\theta &= \theta - \gamma \frac{m_\theta}{\varepsilon+\sqrt{v_\theta}}\label{eq_1.9c}
\end{align}
where $\gamma, \alpha_1,\alpha_2, \varepsilon$ are parameters controlling the learning rate and the convergence of the ANN. The "moments" $v_\theta,m_\theta$ are initialized to zero, while the weights and biases $\theta\equiv \{\hat{W}_{i,j,k}, \beta_{i,j}\}$ are normal random variables with zero average and unit variance. The ADAM algorithm \eqref{eq_1.9a}-\eqref{eq_1.9c} should find, asymptotically, the global minimum of the error function $E$. This means that the ANN should predict output values $Z_{m+1}^p$ as suitable approximations for the real $Y^p, \forall p=\overline{1,N_d}$ and beyond.

Based on numerical experience, the optimal values used for the present ANN are $m = 2$ hidden layers and $n_0=3, n_1=10, n_2=5, n_3=1$ neurons in each layer. The activation functions are $f_{i,j}(z) = \tanh(z), \forall i=\overline{1,m}, j=\overline{1,n_i}$ and $f_{m+1,j}(z) = z^2, \forall j=\overline{1,n_{m+1}}$ for the hidden, respectively, output layers. The parameters for ADAM are $\gamma = 0.001, \alpha_1=0.9,\alpha_2=0.95,\varepsilon=10^{-7}$.

\subsection{Defining the database}
\label{Section_1.4}

At this point, we must recall how the question of CR turbulent transport fits into the formal description of an ANN. The eqns. of the model \eqref{eq_1.1}-\eqref{eq_1.5} have a set of free parameters, namely $\left(T,\mu, b, \lambda_\perp,\lambda_\parallel\right)$. In our simulations, we chose $\mu=0.5$ and $\lambda_\parallel = 10\lambda_\perp$, values relevant to observations from solar winds. The rest, $X = \left(T, b, \lambda_\perp\right)$ represents the input variable from ANN's perspective. For each set of parametric values $X$, associated perpendicular diffusion coefficients are obtained as output $Y=D_\perp$. In order to train and test the ANN, we need to construct a database of $N_d$ pairs $(X^p,Y^p)_{p=1,N_d}$.

The latter is obtained through the test-particle method described in Section \eqref{Section_1.2}. An appropriate number of $N_d = 5\times 10^3$ simulations has been performed. Note that this task is far more demanding than constructing and training the ANN, since it requires a much larger volume of CPU computing time. 

The $n_0-$dimensional space of $X$ parameters must be truncated to a finite domain. This is done by choosing a cube of limiting values: $T\in \left(0, 10\right)$ which for protons corresponds to a kinetic energy $(0eV, 10GeV)$, $b\in \left(0, 1\right)$ corresponding to $(0,B_0)$ amplitude of fluctuations, $\lambda_\perp\in \left(0.2,20\right)$. Note that for a proton, $T=1$ corresponds to $\approx 1GeV$ energy and a scaled Larmor radius $\rho_L\approx 1.7$, thus, comparable with most values of correlation lengths, $\lambda_\perp$.

In order to avoid spurious biases, the values $\{X^p\}$ are generated randomly uniform inside the parametric hypercube. It is important to observe that the database can be completed with analytical results. For example, at $b=0\to D_\perp=D_\parallel = 0$. Approximately $5-10\%$ of the database can be filled with such exact results. If this fraction is too large, the hypercube becomes non-uniformly filled, and the ANN learns predominately the analytical values, which are of no interest. 

From the total database, $90\%$ of the $(X,Y)$ pairs are used for training, while the remaining $10\%$ will serve as testing grounds for the assessment of ANN's predictive abilities.

\section{Results}
\label{Section_2}

\subsection{Turbulent transport}
\label{Section_2.1}

The transport model has been used in the framework of the test-particle method to evaluate CR trajectories and, consequently, perpendicular diffusion coefficients $Y=D_\perp$ for different random combinations of free input parameters $X\equiv\left(T,b,\lambda_\perp\right)$ values. The equations of motion \eqref{eq_1.1} are fully relativistic and do not rely on the guiding-center approximation \cite{Hauff_2010}. Consequently, particle trajectories describe the full Larmor rotation induced by the average magnetic field $B_0$ as well as the "scattering" in the fluctuating field $\mathbf{b}$ which is 2D. This level of description is needed whenever the characteristic length scales of fluctuations, in our case $\lambda_\perp$, are comparable with the Larmor radius $\rho_L$. This is precisely the case for the database since for most energies $\rho_L\sim 1$ (in the scaling described in Section \eqref{Section_1.1}) while $\lambda_\perp\in\left(0.2,20\right)$.
\begin{figure}
	\subfloat[\label{fig_1_a}]{
		\includegraphics[width=.88\linewidth]{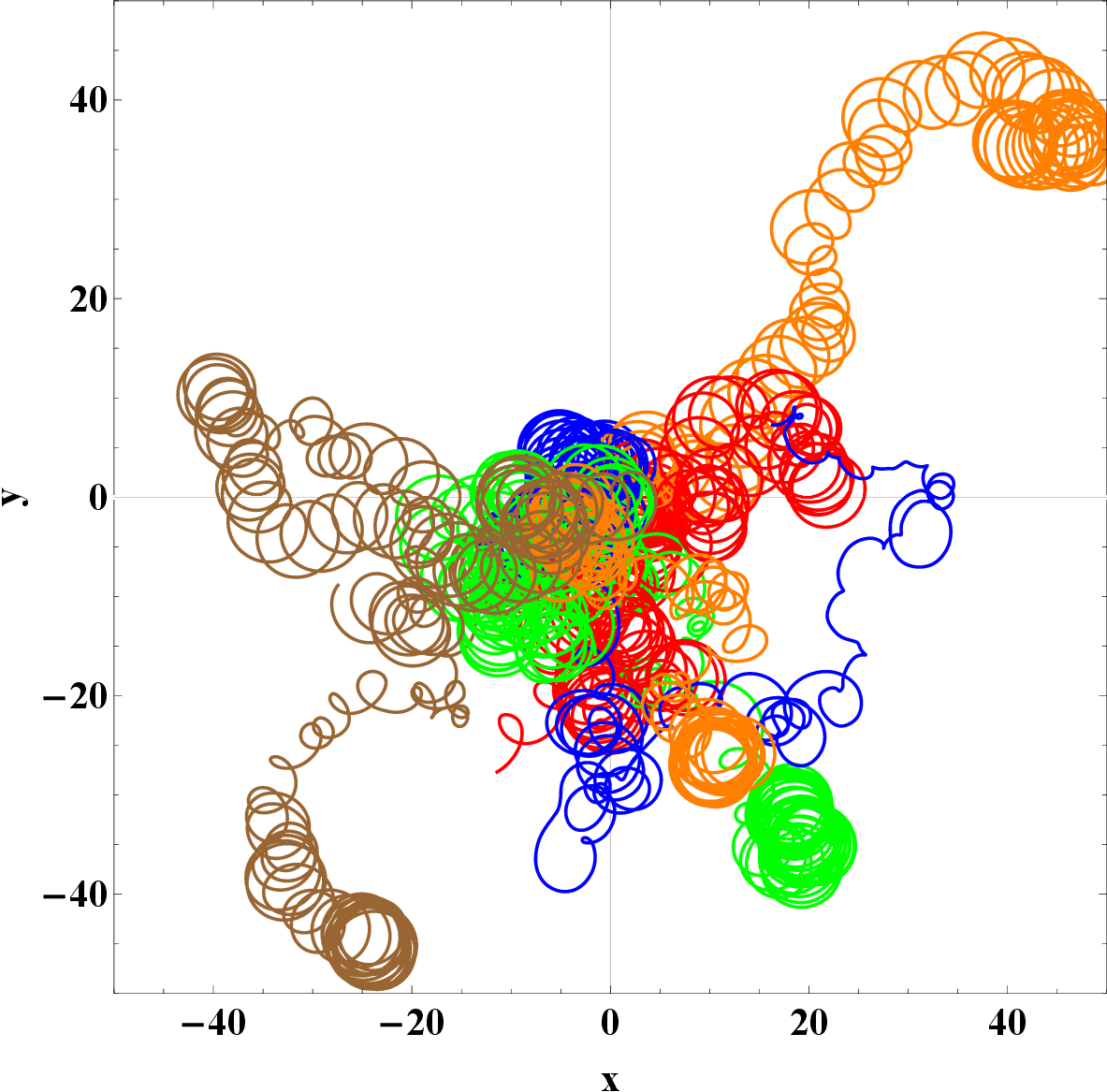}%
	}\\
	\subfloat[\label{fig_1_b}]{
		\includegraphics[width=.88\linewidth]{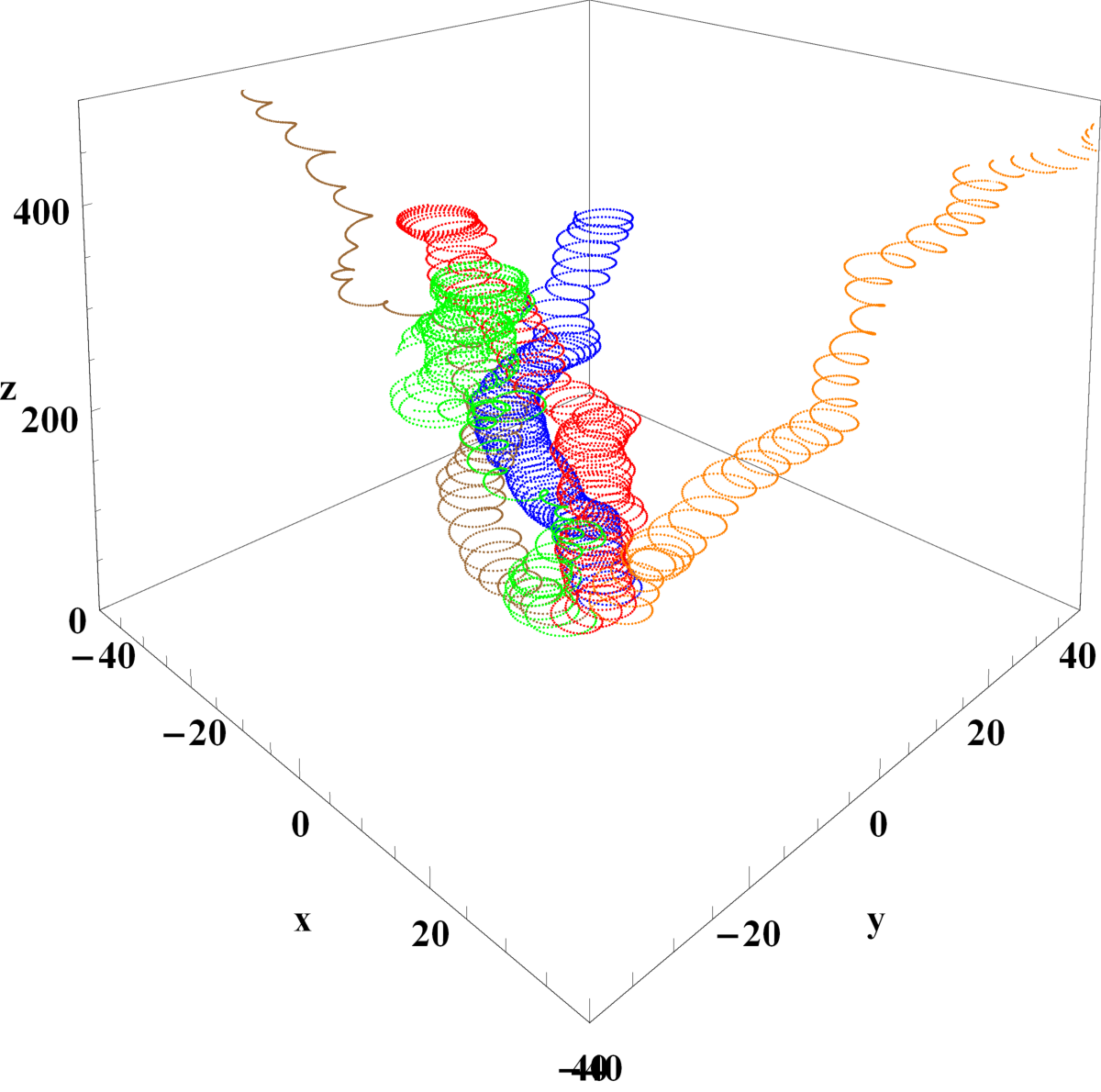}%
	}
	\caption{Typical CR trajectories in turbulent 2D magnetic fields shown in the perpendicular plane (a) and in full 3D geometry (b).}
\end{figure}

In Figs. \eqref{fig_1_a}-\eqref{fig_1_b} are shown typical CR trajectories in the perpendicular plane (Fig. \eqref{fig_1_a}) and in full 3D space (Fig. \eqref{fig_1_a}). Due to the fact that the pitch angle has been set to $\mu=0.5$ in all simulations, the particles propagate along the parallel direction with larger velocities than the perpendicular guinding-center drifts. The Larmor rotation and the scattering due to fluctuations are obvious.

\begin{figure}
	\includegraphics[width=.93\linewidth]{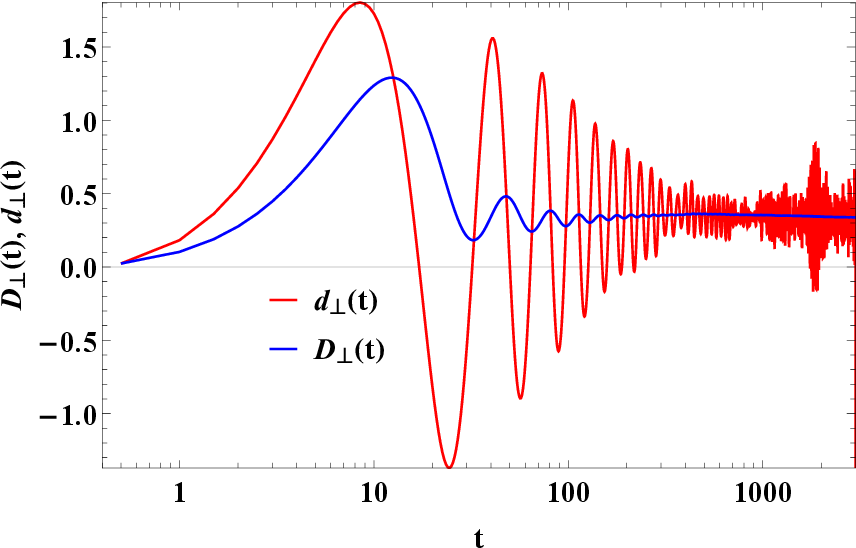}%
	\caption{Typical perpendicular diffusion coefficients computed in two ways.}
	\label{fig_1_c}
\end{figure}
	
The running diffusion coefficient $D_\perp(t)$ related to the mean-square-displacement of particles, $\langle x^2(t)\rangle$, is computed accordingly with eq. \eqref{eq_1.5}. Another, more frequently used, formula for diffusion is $d(t)=\partial_t\langle x^2(t) \rangle/2$. The reason for which $D_{\perp}(t)$ is used here instead of $d_\perp(t)$ is due to the fact that we capture Larmor rotations in our model. This component of motion exhibits large fluctuations at small times in $d_\perp(t)$. $D_\perp(t)$, on the other hand, due to its functional form, suppresses the effect of rotations very fast as well as the statistical fluctuations at long times. Nonetheless, both forms converge asymptotically to the same value. All these features can be seen in Fig. \eqref{fig_1_c} where two typical time dependent diffusion profiles are shown for the same set of free parameters.

It must be emphasized that our database describes only pure diffusive regimes (in the perpendicular plane). In general, anomalous behaviour is possible as it was described, many times, in literature \cite{BHATTACHARJEE2000109,Beresnyak_2011,Giacalone_1999,Hauff_2010,POMMOIS2005647}. In fact, the finite parallel correlation length $\lambda_\parallel = 10\lambda_\perp$ is responsible for the decay of the Lagrangian correlation and, consequently, the saturation of the running diffusion to a constant value, thus, diffusive transport. The mechanism is simple: the parallel motion of CRs in fields with parallel dependence induces an effective decorrelation time $\tau_c \approx \lambda_\parallel/v_\parallel$. Describing subdiffusive or superdiffusive transport remains an important task for future developments of ANNs.

\subsection{Convergence properties of the ANN}
\label{Section_2.2}

Once the database has been build and the ANN constructed (programmed), we are all set to start the training phase. A random configuration of initial weights and biases $\{W_{i,j,k},\beta_i\}$ is chosen and the minimization procedure (via ADAM algorithm) begins. The random nature of initial $\{W_{i,j,k},\beta_i\}$ is needed to avoid setting the ANN in a configuration point from which won't be able to converge to the global minimum. Due to the same reason, different initializations lead to distinct paths of convergence. There is also a question of weather the true global minimum is achieved \cite{doi:10.1126/science.aaa8415} or the algorithm gets stuck a neighbouring local minimum. Regardless, the practical experience has shown that the asymptotic states are appropriate approximations of the real minimum.

\begin{figure}
	\includegraphics[width=.93\linewidth]{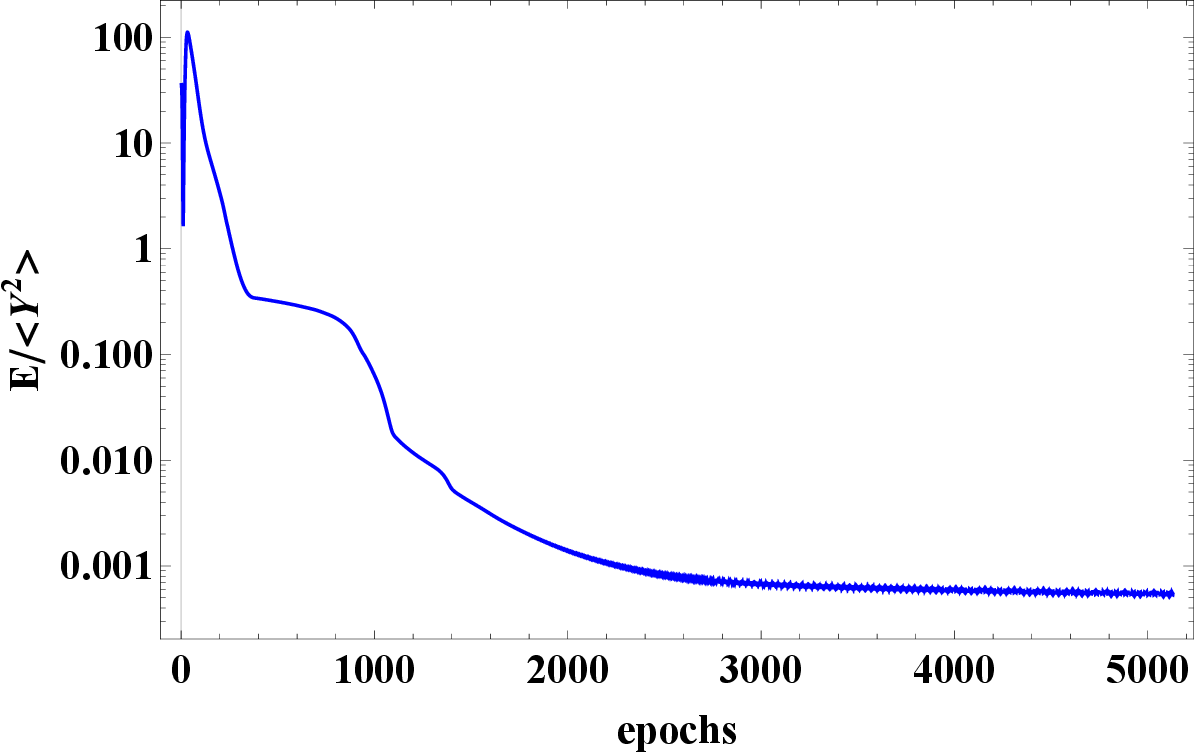}%
	\caption{Error $E$ evolution during the iterative training of the ANN with ADAM.}
	\label{fig_2_a}
\end{figure}

In Fig. \eqref{fig_2_a} it is shown a typical evolution of the error \eqref{eq_1.8} relative to the norm of output values. The learning stage is fast, approximately $5 s$ on a typical personal CPU. One must note that the error function is almost always decreasing, but the evolution can manifest periods of quasi-plateau, as the one between $400-900$ epochs in Fig. \eqref{fig_2_a}. For this reason, it is important to allow for long periods of minimization, in order to avoid confusing a local minima (the plateau) with the global one.

The number of particles propagated within a single simulation in this work is set to $N_p =10^4$. Looking at Fig. \eqref{fig_1_c}, it might seem that this value is unnecesary since the diffusion coefficient $D_\perp(t)$ saturates to a nice, constant, plateau without temporal fluctuations. While this is true, this does not mean that the ensemble is sufficiently well represented and, consequently, that the asymptotic value is statistically robust (without fluctuations). In other words, if $N_p$ is too small, for two identical simulations we might obtain relatively different diffusions. This would have a detrimental impact on the database, introducing a spurious numerical fluctuation and making the job of convergence much more difficult.

\begin{figure}
	\includegraphics[width=.93\linewidth]{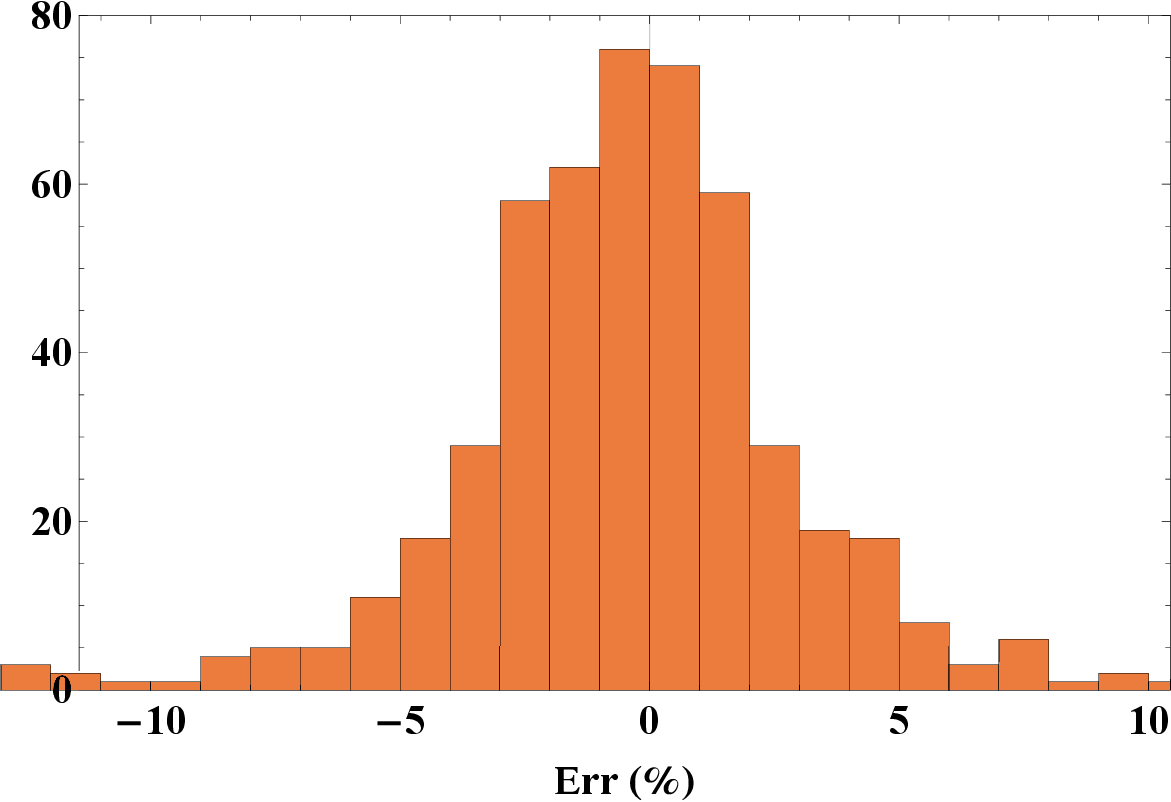}%
	\caption{Relative error distribution on the testing database.}
	\label{fig_2_b}
\end{figure}

When the ADAM algorithm reaches the final plateau and an acceptable small error, one is ready to enter the testing phase. Now, the ANN, that is to say the values of $\{W_{i,j,k},\beta_i\}$ obtained after minimization, can be used to evaluate diffusion coefficients for the remaining $10\%$ input, $X^{test}$, of the database that was not used in the training stage. The predicted values, $Z_{m+1}^{test}$, are compared with exact output diffusions $Y^{test}$ from the database. Fig. \eqref{fig_2_b} shows the histogram of relative errors between ANN's predictions and real output, $Err = 100\left(Z_{m+1}^{test}/Y^{test}-1\right)$.

While there are points with relatively large errors (usually those of very small diffusion $Y^{test}\ll 1$), the overall accuracy of the ANN can be estimated with the second moment of the histogram and it is close to $5\%$. 

When comparing computing times between test-particle simulations, $t_{sims}$, and ANN, $t_{ANN}$, the latter shows its true power: for a single diffusion coefficient on a two-processor CPU $t_{sim} \approx 2s$ while $t_{ANN}\approx 10^{-7}s$. Thus, ANNs are $10^7$ times faster than numerical simulations.

\subsection{Making predictions with the ANN}
\label{Section_2.3}

\begin{figure}
	\subfloat[\label{fig_3_a}]{
		\includegraphics[width=.88\linewidth]{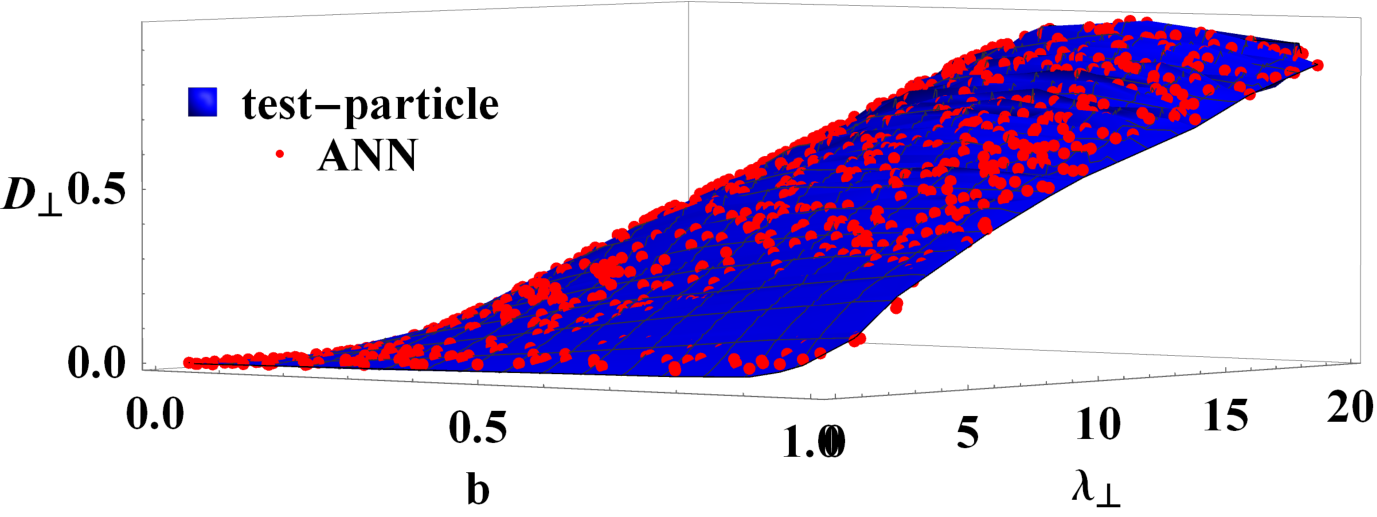}%
	}\\
	\subfloat[\label{fig_3_b}]{
		\includegraphics[width=.88\linewidth]{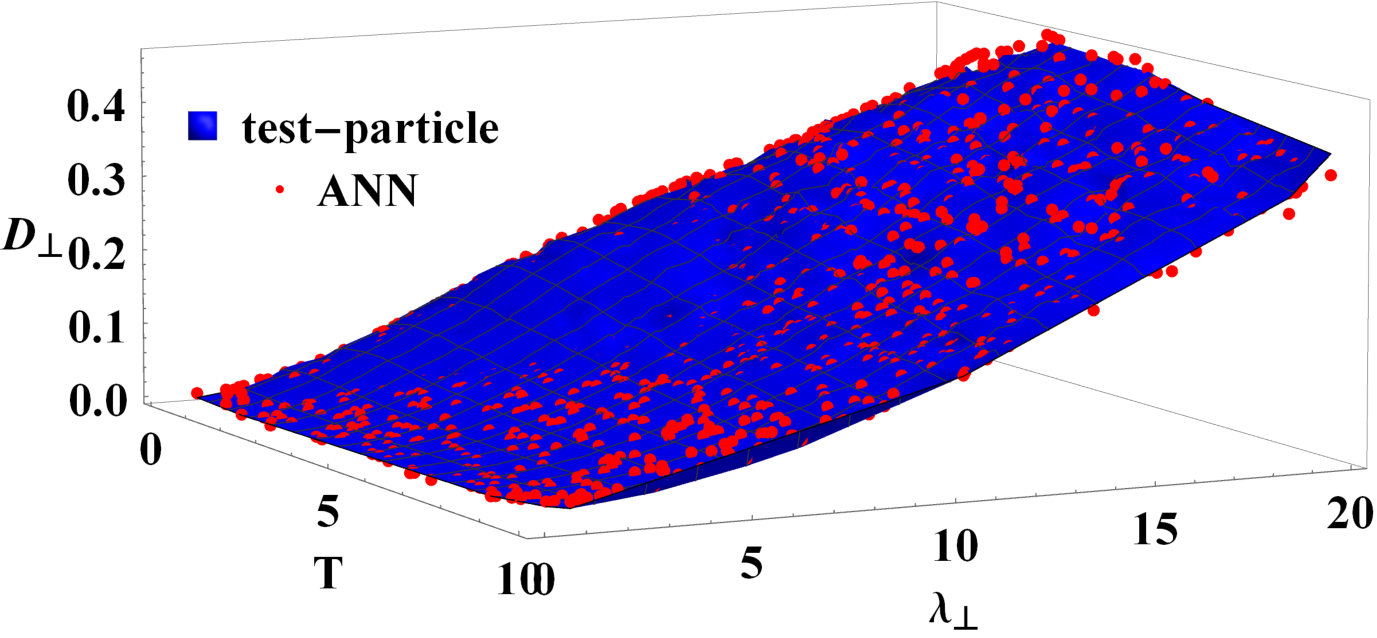}%
	}\\
	\subfloat[\label{fig_3_c}]{
		\includegraphics[width=.88\linewidth]{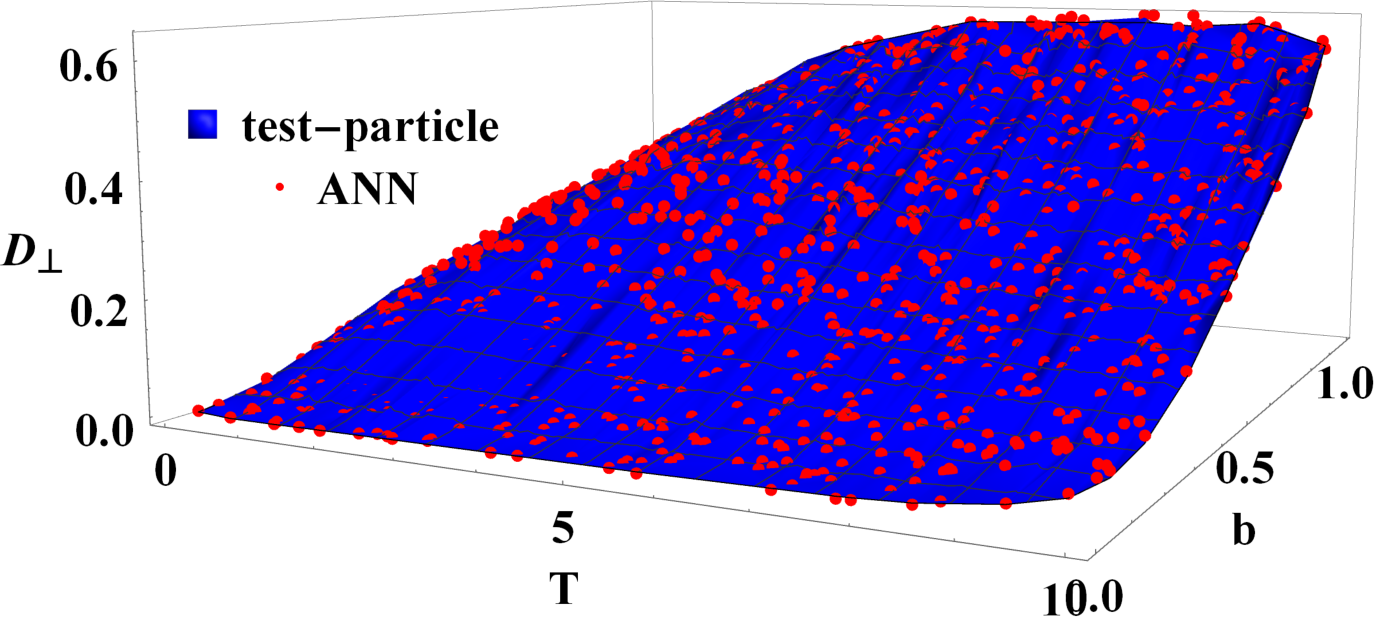}%
	}
	\caption{ANN predicted diffusion (red dots) in comparison with exact, test-particle derived, data (blue surface) keeping one parameter constant at a time ($T=1$-(a), $b=0.2$-(b), $\lambda_\perp=5$-(c)).}
\end{figure}

The histogram of errors obtained in the testing phase \eqref{fig_2_b} suggests a global error of $\sim 5\%$. But this is not enough to ensure an acceptable accuracy of the ANN. In fact, there are points which have departed more than $10\%$ from the exact values. Moreover, these errors might be a serious liability if they are clustered in some special way inside the parametric space of input variables.

In order to test whether this is the case, we must look for reduced spaces inside the database and evaluate if the ANN's predictions are a good fit for the real data. We do that on two distinct levels: first we look at two-dimensional dependencies of $D_\perp$ varying two parameters at a time, than, we look at simple dependencies between $D_\perp$ and only one parameter. 

\begin{figure}
	\subfloat[\label{fig_4_a}]{
		\includegraphics[width=.88\linewidth]{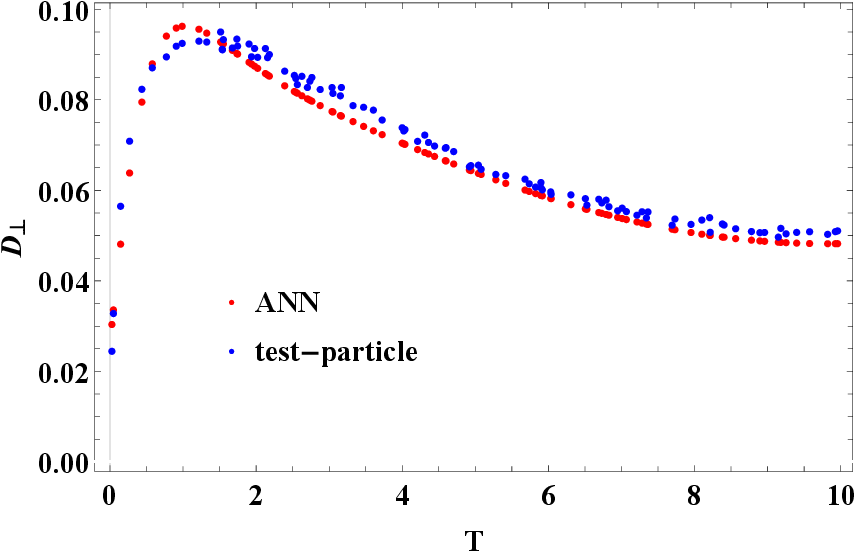}%
	}\\
	\subfloat[\label{fig_4_b}]{
		\includegraphics[width=.88\linewidth]{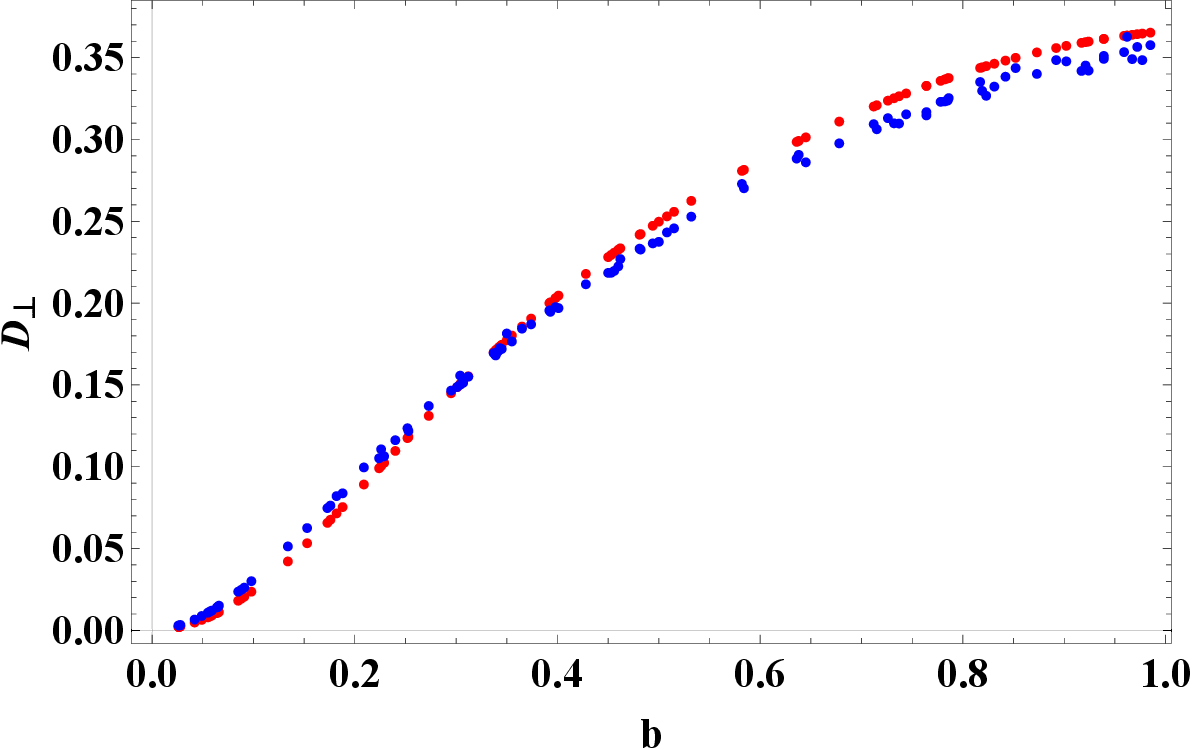}%
	}\\
	\subfloat[\label{fig_4_c}]{
		\includegraphics[width=.88\linewidth]{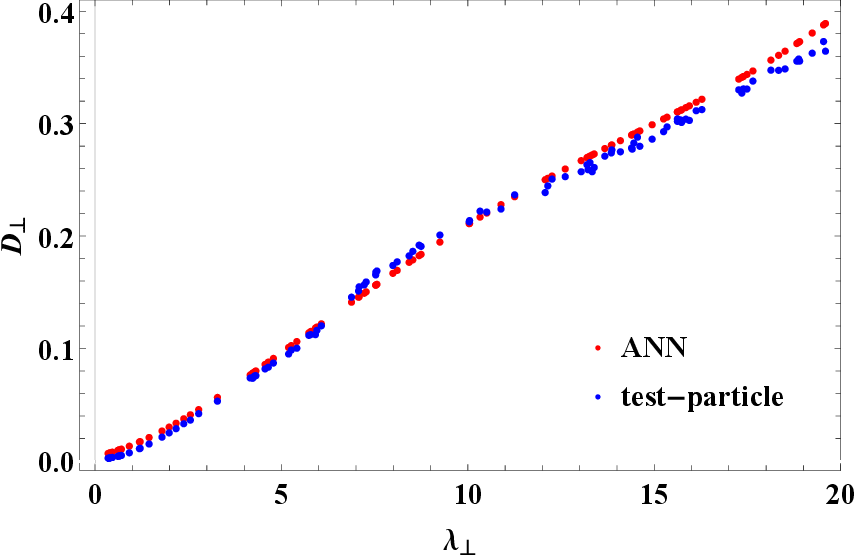}%
	}
	\caption{ANN predicted diffusion (red dots) in comparison with exact, test-particle derived, data (blue dots) keeping two parameters constant at a time ($b=0.2, \lambda_\perp=5$-(a), $T=1, \lambda_\perp=5$-(b), $T=1, b=0.2$-(c)).}
\end{figure}

In Figs. \eqref{fig_3_a}-\eqref{fig_3_c} we compare ANN output (red dots) with exact values (blue surface, obtained with test-particle simulation). Fig. \eqref{fig_3_a} uses as variables the magnetic turbulence amplitude $b$ and the perpendicular length $\lambda_\perp$ while the energy is set to $T=1$. Fig. \eqref{fig_3_b} sets $b=0.2$ and varies $\lambda_\perp$ and $T$, while in Fig. \eqref{fig_3_c} $\lambda_\perp = 5$. As one can see, all predicted values (red) lie close to the surface of exact diffusions (blue) in all cases. Thus, there is no special parametric domain where errors tend to accumulate and the ANN exhibits similar accuracy across the entire database. 

In Figs.  \eqref{fig_4_a}-\eqref{fig_4_c} we set two parameters to constant values and vary only the third, as it follows: Fig. \eqref{fig_4_a} uses $b=0.2, \lambda_\perp=5$, Fig. \eqref{fig_4_b} $T=1, \lambda_\perp=5$ and Fig. \eqref{fig_4_c} $T=1, b=0.2$. The overlap between predictions (red) and exact data (blue) are, again, consistently accurate across all simulations. Furthermore, at the level of such single parameter dependencies, we can understand how using an ANN tool could enhance our ability to investigate and understand the physical processes at play. For example, having easy access to the red curve in Fig. \eqref{fig_4_a} allows one to observe the maxima in the diffusion profile and infer about the existence of two competing mechanisms in the influence of particle energy. One is related to the monotonic increase of parallel velocity $v_\parallel$ with $T$ which increases the transport, while the other is connected to the finite Larmor radius effects that tend to decrease diffusion \cite{Hauff_2010}.

\section{Conclusions}
\label{Section_3}

The present work described a methodology for building artificial neural networks designed for predictions of cosmic ray transport in turbulent magnetic fields. The ANN developed here has a standard architecture with two hidden layers and $tanh$/quadratic activation functions. It uses the ADAM algorithm for optimizations in the learning phase. The input data in the training/testing database consists of the values of free parameters for CRs and turbulence, chosen inside a (hyper)cube of convenience. The output is represented by associated diffusion coefficients. The values of the latter are evaluated using a transport model which is numerically tackled with the aid of test-particle simulations. 

The learning stage showed fast and good convergence properties. In the testing phase, a good fit between exact and ANN predicted data was found with overall errors of $\approx 5\%$. The predictive power of the ANN is proven by the dependencies of the transport coefficients on individual parameters. The most stringent feature of the network is its speed, since it is able to compute diffusion on bulk cases approximately $10^7$ times faster than test-particle simulations.

One might argue that the overall database constructed here is too limited: the model of turbulent fields (2D with a specific spectrum and connected correlation lengths $\lambda_\parallel = 10\lambda_\perp$) is too simple to represent most astrophysical regimes and the numerical range of parameters (energy, correlation lengths, field strengths) is quite narrow. This is all true, but the purpose of this work was not to exhaust wide physical regimes. It was the author's intention to present a proof of concept for the use of machine learning techniques in the field of astrophysics, applied to a particular problem.

Hopefully, the community will explore the use of this tool in future works using more sophisticated transport models and wider databases, thus, developing more potent ANNs. It is likely that, given the numerical effort required in building large databases, collaborations will be suitable.

\emph{A word of caution.} Developing ANNs to evaluate quantities of interest is not real knowledge. It does not constitute insight into any physical processes or quantitative mechanisms. Nonetheless, for practical applications where predictions are important, ANNs might be the  choice of preference. Finally, even for more academically oriented questions, having a fast tool at hand to explore different regimes could be invaluable in the quest for disentangling the physical mechanisms at work.

\section*{Statement}

The database and the ANN developed in this study are available upon reasonable request from the authors.

\section*{Acknowledgements}
This research was partially supported by Romanian Ministry of Research, Innovation and Digitalization under Romanian National Core Program LAPLAS VII – contract no. 30N/2023.

\bibliographystyle{apsrev}
\bibliography{biblio}

\end{document}